\documentclass[twocolumn,showpacs,preprintnumbers,amsmath,amssymb,prl]{revtex4-1}

\usepackage{graphicx}
\usepackage{dcolumn}
\usepackage{bm}
\usepackage{psfrag}

\begin{document}

\preprint{APS/123-QED}

\title{Transitions in rapidly rotating convection driven dynamos}

\author{A. Tilgner}

\affiliation{Institute of Geophysics, University of G\"ottingen,
Friedrich-Hund-Platz 1, 37077 G\"ottingen, Germany }

\date{\today}

\begin{abstract}
Numerical simulations of dynamos in rotating Rayleigh-B\'enard convection in
plane layers are presented. Two different types of dynamos exist which obey different
scaling laws for the amplitude of the magnetic field. The transition between the
two occurs within a hydrodynamically uniform regime which can be classified as
rapidly rotating convection.
\end{abstract}

\pacs{91.25.Cw, 47.65.-d}
\maketitle

Most celestial bodies are the seat of a magnetic field. It is commonly assumed
that this field is either a fossil field remaining from the formation process of the
body, or that it is maintained by the dynamo effect which converts kinetic
energy of the motion of a liquid or gaseous electric conductor into magnetic
energy. It is in many cases reasonable to suppose that the motion of the fluid
conductor is driven by convection. The minimal model of a convection driven
dynamo is a horizontal plane layer in a gravity field,
filled with electrically conducting fluid,
heated from below and cooled from above, and rotating about a vertical axis.
Such a plane layer may be viewed as a local approximation to the astrophysically
more relevant spherical geometry.

Several studies of this problem exist and they have found different
types of magnetic fields \cite{Jones00,Rotvig02,Stellm04,Cattan06}.
Depending on whether rapidly or
slowly rotating layers are simulated, the magnetic fields are either generated
at the largest available spatial scale or at smaller scales. The generated
magnetic fields contain either a large or small mean field (where the mean is
obtained by averaging over horizontal planes), and the magnetic field in the
final statistically stationary state either executes small fluctuations around a
well defined mean or experiences large swings during which the field amplitude
varies by an order of magnitude and more. 

This paper presents simulations of dynamos in rotating and convecting plane
layers. If the convection is driven stronger
and stronger at fixed rotation rate, the flow behaves at some point as if
it was not rotating. This transition shows in the scaling of the heat transport
which can be used to distinguish slow from rapid rotation \cite{Schmit09, King09}.
Within the convection flows which are rapidly rotating according to this criterion,
it will be shown below that different types of dynamos exist.

Consider a plane layer with boundaries perpendicular to the $z-$axis, rotating
about this axis with angular velocity $\bm \Omega = \Omega \hat{\bm z}$, where
the hat denotes a unit vector. Gravity $\bm g$ is pointing along the negative
$z-$direction, $\bm g = -g \hat{\bm z}$. The layer is filled with 
electrically conducting fluid of density $\rho$, kinematic viscosity $\nu$,
thermal diffusivity $\kappa$, thermal expansion coefficient $\alpha$, and
magnetic diffusivity $\lambda$. The fluid is moving according to the velocity
field $\bm v(\bm r,t)$ which depends on position $\bm r$ and time $t$. The fluid
is also permeated by the magnetic field $\bm B(\bm r,t)$ and has temperature
$T(\bm r,t)$. 

The boundaries are located in the planes $z=0$ and $z=d$. These boundaries are
assumed to be at constant temperatures $T_0+\Delta T$ and $T_0$, respectively.
The boundaries are also assumed to be free slip and the space outside the fluid
layer is assumed to be a perfect conductor. The boundary conditions therefore
read
$T=T_0+\Delta T$ at $z=0$, $T=T_0$ at $z=d$, and furthermore
$v_z = \partial_z v_y = \partial_z v_x =
B_z = \partial_z B_y = \partial_z B_x =0$ at $z=0$ and $d$.
Periodic boundary conditions are imposed in the lateral directions forcing all
fields to have periodicity lengths $l_x$ and $l_y$ in the $x-$ and
$y-$directions.

The equations have been solved numerically in the nondimensional form which
obtains if one uses $d$, $d^2/\kappa$, $\kappa/d$, $\rho \kappa^2/d^2$,
$\Delta T$ and $\sqrt{\mu_0 \rho} \kappa/d$ as units of length, time, velocity,
pressure, temperature difference from $T_0$, and magnetic field, respectively. 
Four control parameters appear in the equations: The Rayleigh number
$\mathrm{Ra}$, the Ekman number $\mathrm{Ek}$, the Prandtl number $\mathrm{Pr}$,
and the magnetic Prandtl number $\mathrm{Pm}$. They are defined by
\begin{equation}
\mathrm{Ra}=\frac{g \alpha \Delta T d^3}{\kappa \nu} ~~~,~~~
\mathrm{Ek}=\frac{\nu}{\Omega d^2} ~~~,~~~
\mathrm{Pr}=\frac{\nu}{\kappa} ~~~,~~~
\mathrm{Pm}=\frac{\nu}{\lambda}
\end{equation}

The nondimensional equations read (using the same variables for the nondimensional
quantities as their dimensional counter parts):
\begin{equation}
\partial_t \rho + \nabla \cdot \bm v = 0
\label{eq:conti_BQ}
\end{equation}
\begin{equation}
\begin{split}
\partial_t \bm v +(\bm v \cdot \nabla) \bm v
+ 2 \frac{\mathrm{Pr}}{\mathrm{Ek}} \hat{\bm z} \times \bm v = \\
- c^2 \nabla \rho +
\mathrm{Pr} ~ \mathrm{Ra} ~\theta \hat{\bm z} + \mathrm{Pr} \nabla^2 \bm v
+ (\nabla \times \bm B) \times \bm B
\end{split}
\label{eq:NS_BQ}
\end{equation}
\begin{equation}
\partial_t \theta + \bm v \cdot \nabla \theta -v_z =\nabla^2 \theta
\label{eq:T_BQ}
\end{equation}
\begin{equation}
\partial_t \bm B +\nabla \times (\bm B \times \bm v) =
\frac{\mathrm{Pr}}{\mathrm{Pm}} \nabla^2 \bm B
\label{eq:induc_BQ}
\end{equation}
\begin{equation}
\nabla \cdot \bm B = 0
\label{eq:div_BQ}
\end{equation}
where $\theta$ is the deviation from the conductive temperature profile so that
$\theta=0$ at the top and bottom boundaries. From now on, only nondimensional
variables will be used.

The above equations approximate the Boussinesq equations using the method of
artificial compressibility \cite{Chorin67}, which assumes an equation of state
of the form $p=c^2 \rho$, where $p$ stands for the pressure and $c$ for the
velocity of sound. If one develops the variables $p$, $\bm v$, $\theta$ and $\bm
B$ in powers of $1/c^2$ (as for example in
$\theta=\theta_0 + \frac{1}{c^2} \theta_1 + \frac{1}{c^4} \theta_2 + ...$) and
inserts these series into equations (\ref{eq:conti_BQ}-\ref{eq:div_BQ}) together
with $p=c^2 \rho$, one recovers the Boussinesq equations at the order 
$(1/c^2)^0$. This method simulates the standard Boussinesq
equations if the Mach number is small and if the time it
takes sound to travel across the layer is much less than the rotation period,
i.e. if $c^2 \gg (1/2 \pi)^2 (\mathrm{Pr}/\mathrm{Ek})^2$. In all the
simulations reported here, the Mach number was less than 0.1 and
$c^2 \mathrm{Ek}^2 > 0.4$. The validity of the entire approach was tested by
reproducing a few results of ref. \cite{Stellm04,Schmit09} in which the Boussinesq
equations are solved.
The numerical simulations used a
finite difference method implemented on GPUs. The method is the same as the one
in ref. \cite{Tanriv11, *Tilgne11} except that fourth order spatial derivatives
have been used in the induction and all diffusion terms. Spatial resolutions
have been adapted to the control parameters and went up to $256^3$. The runs
were divided into six series, three for $\mathrm{Pm}=1$ and the other three for
$\mathrm{Pm}=3$, while $\mathrm{Pr}$ was always set to $0.7$. For each
$\mathrm{Pm}$, the Ekman numbers $2 \times 10^{-4}$, $2 \times 10^{-5}$, and $2
\times 10^{-6}$ have been used. Within each of the six series, $\mathrm{Ra}$ was
varied from its critical value up to 100 times critical for $\mathrm{Ek}=2
\times 10^{-4}$ and three times critical for $\mathrm{Ek}=2
\times 10^{-6}$. Because the typical length scale of convection varies with
$\mathrm{Ek}$, the periodicity lengths were set to $l_x/d = l_y/d =1$, $1/2$,
and $1/4$ for $\mathrm{Ek}=2 \times 10^{-4}$, $2 \times 10^{-5}$ and 
$2 \times 10^{-6}$, respectively. All runs have been started from magnetic seed
fields small enough so that the seed field does not modify the convection and
cannot lead to subcritical bifurcation.

The output of the computations include the densities of kinetic and magnetic
energies, $e_{\mathrm{kin}}$ and $e_B$, defined by
\begin{equation}
e_{\mathrm{kin}}= \frac{1}{V} \int \frac{1}{2} \bm v^2 dV ~~~,~~~
e_B= \frac{1}{V} \int \frac{1}{2} \bm B^2 dV,
\end{equation}
where the integration extends over the entire fluid volume $V$. If we denote the
time average by angular brackets, one can compute average energy densities
$E_{\mathrm{kin}}$ and $E_B$ from $E_{\mathrm{kin}}=\langle e_{\mathrm{kin}}
\rangle$ and $E_B=\langle e_B \rangle$ as well as the Reynolds number 
$\mathrm{Re}$ from $\mathrm{Re}=\langle \sqrt{2 e_{\mathrm{kin}}} \rangle /
\mathrm{Pr}$. The magnetic Reynolds number $\mathrm{Rm}$ is defined as 
$\mathrm{Rm}=\mathrm{Re} \, \mathrm{Pm}$.

\begin{figure}
\includegraphics[width=8cm]{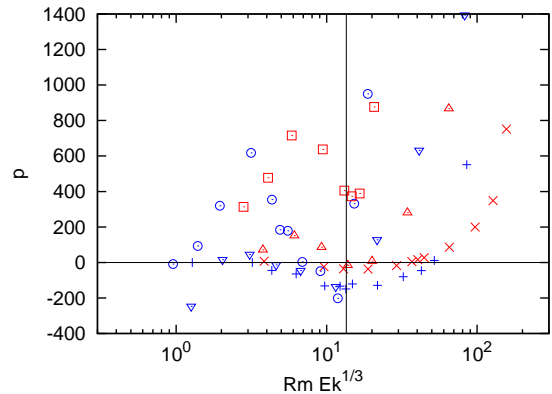}
\caption{(Color online)
Growth rate $p$ of magnetic energy in kinematic dynamo calculations as a
function of $\mathrm{Rm}\, \mathrm{Ek}^{1/3}$. Results for $Pm=1$ are shown
in blue and those for $Pm=3$ are in red. For $Pm=1$, the Ekman numbers of
$2 \times 10^{-4}$, $2 \times 10^{-5}$, and $2 \times 10^{-6}$ are indicated by
the plus sign, triangle down, and circle, respectively, whereas for $Pm=3$, the
same Ekman numbers are indicated by the x sign, triangle up, and square. The
vertical line is located at $\mathrm{Rm}\, \mathrm{Ek}^{1/3}=13.5$.}
\label{fig:kinematic}
\end{figure}

When the Rayleigh number is increased within each series of computations at
constant $\mathrm{Ek}$ and $\mathrm{Pm}$, one finds an onset of dynamo action,
and in most series an interval of $\mathrm{Ra}$ in which self-sustained magnetic
fields do not exist before dynamo action starts anew at higher $\mathrm{Ra}$. A
preliminary set of computations solved the kinematic dynamo problem, in which
the back reaction of the magnetic field on the flow is neglected. For this
purpose, the $(\nabla \times \bm B) \times \bm B$ term was removed form eq.
(\ref{eq:NS_BQ}) and the remaining system of partial differential equations
integrated in time. Without the $(\nabla \times \bm B) \times \bm B$ term,
initial magnetic seed fields either grow or decay indefinitely after transients
have passed. They do so in a fluctuating manner if the flow is chaotic, but
viewed on long time scales, the magnetic energy grows or decays exponentially as
$e^{p t}$. The growth rate $p$ is shown as a function of $\mathrm{Rm}\,
\mathrm{Ek}^{1/3}$ in fig. \ref{fig:kinematic}. In this representation, the
local minimum of $p$ is found at the same position for all $\mathrm{Ek}$ and
$\mathrm{Pm}$ near $\mathrm{Rm}\, \mathrm{Ek}^{1/3}=13.5$. Choosing 
$\mathrm{Rm}\, \mathrm{Ek}^{1/3}$ as parameter does not lead to an overlap of
all data in fig. \ref{fig:kinematic}. For instance, the local maximum of $p$ at
$\mathrm{Rm}\, \mathrm{Ek}^{1/3}<13.5$ occurs at various values of
$\mathrm{Rm}\, \mathrm{Ek}^{1/3}$ and analytical work tells us that the dynamo
onset (which was not accurately located in the present simulations) occurs at
$\mathrm{Rm}\, \mathrm{Ek}^{1/6}=const.$ \cite{Soward74}. The product
$\mathrm{Rm}\, \mathrm{Ek}^{1/3}$ is proportional to the magnetic Reynolds
number on the length scale of one column in the convective flow, because
in the limit of
small $\mathrm{Ek}$, the wavelength of the flow at the onset of convection is
given by $3.83 \cdot \mathrm{Ek}^{1/3}$. At the local minimum, $p$ is negative
in all cases except for $\mathrm{Ek}=2\times 10^{-6}$, $\mathrm{Pm}=3$. Since
$\mathrm{Rm}$ increases monotonically with $\mathrm{Ra}$ (see fig.
\ref{fig:phase_diagram} below), this means that a stronger driving does not
necessarily help the dynamo.

This conclusion is also valid for no slip boundary conditions. Ekman layers
appear at no slip boundaries which induce vertical flow in the bulk and
potentially help the dynamo effect by increasing helicity. However, the same
sequence of positive and negative growth rates shown in fig. \ref{fig:kinematic}
was also found for no slip boundaries by varying $\mathrm{Ra}$ for
$\mathrm{Ek}=2\times 10^{-4}$, $\mathrm{Pm}=3$.

From now on, all results are for the full set of equations
(\ref{eq:conti_BQ}-\ref{eq:div_BQ}) including the $(\nabla \times \bm B) \times
\bm B$ term. A global view of all simulations is presented in fig.
\ref{fig:phase_diagram}. The successful and failed dynamos are shown in the plane
spanned by $\mathrm{Rm}\, \mathrm{Ek}^{1/3}$, the parameter already used in
fig. \ref{fig:kinematic}, and $\mathrm{Re}\,
\mathrm{Pr}\, \mathrm{Ek}^{1/2}$. This parameter was introduced in
\cite{Schmit09, Schmit10} because in nonmagnetic convection, the nondimensional
heat flux, expressed as the Nusselt number $\mathrm{Nu}$ and combined with 
$\mathrm{Ek}$ into the quantity $\mathrm{Nu} \, \mathrm{Ek}^{1/3}$,
obeys simple scaling laws in the
two limits of $\mathrm{Re}\, \mathrm{Pr}\, \mathrm{Ek}^{1/2}$ small
(corresponding to flows dominated by rotation) and large (approaching the limit
of nonrotating flows). The two asymptotes fitting the two limits cross at 
$\mathrm{Re}\, \mathrm{Pr}\, \mathrm{Ek}^{1/2}=2$, which therefore separates the
flows dominated by the Coriolis force from those who behave the same
as nonrotating flows in as far as the scaling of the heat flux is concerned. (It
remains to be investigated how this criterion relates to changes in flow
structures or statistics of velocity fluctuations
\cite{Steven10, *Kunnen06}.) It is seen in fig. \ref{fig:phase_diagram} that the transition
line at $\mathrm{Rm}\, \mathrm{Ek}^{1/3}=13.5$ identified in fig.
\ref{fig:kinematic} is crossed at $\mathrm{Re}\, \mathrm{Pr}\,
\mathrm{Ek}^{1/2}<2$ by all series except the one for $\mathrm{Ek}=2 \times
10^{-4}$, $\mathrm{Pm}=1$.


\begin{figure}
\includegraphics[width=8cm]{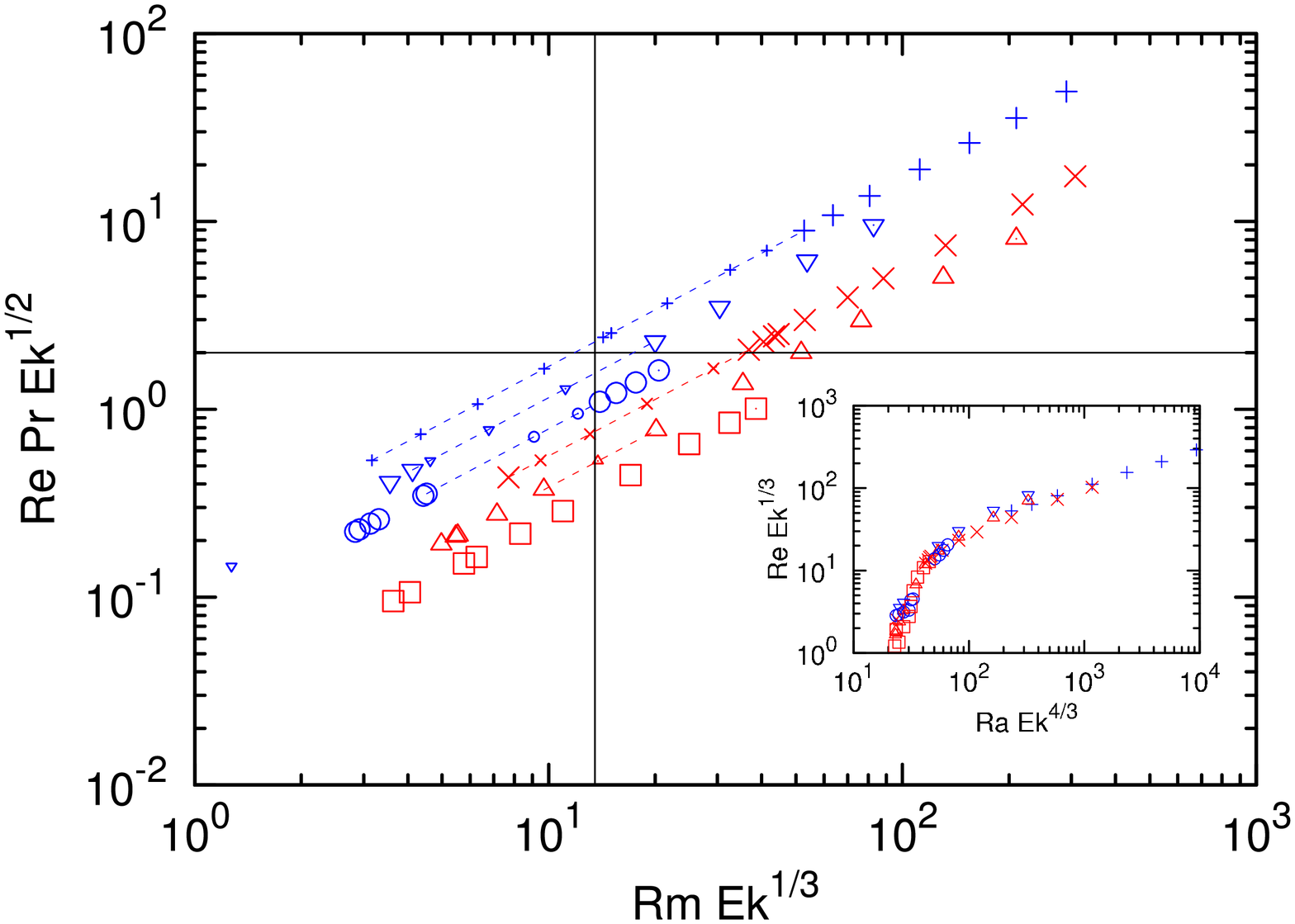}
\caption{(Color online)
Successful dynamos are shown with large and failed dynamos with small symbols
(located on a broken line in the window without dynamo action) in
the plane spanned by $\mathrm{Rm}\, \mathrm{Ek}^{1/3}$ and $\mathrm{Re}\,
\mathrm{Pr}\, \mathrm{Ek}^{1/2}$. The meaning of the symbol shapes is the same
as in fig. \ref{fig:kinematic} and the lines indicate
$\mathrm{Rm}\, \mathrm{Ek}^{1/3}=13.5$ and $\mathrm{Re}\,
\mathrm{Pr}\, \mathrm{Ek}^{1/2}=2$. The inset shows $\mathrm{Re}\,
\mathrm{Ek}^{1/3}$ (which is proportional to the Reynolds number based on the
size of the convection columns) as a
function of $\mathrm{Ra}\, \mathrm{Ek}^{4/3}$ (which is proportional to
the ratio of $\mathrm{Ra}$ to the critical Rayleigh number for the onset of
convection) for the same simulations.}
\label{fig:phase_diagram}
\end{figure}

It will now
be argued that the dynamos both below and above the transition are related to
dynamos known from previous work. Below the transition, the dynamos qualify as
$\alpha^2-$dynamos in the language of mean-field magnetohydrodynamics and rely
on the helicity of the flow \cite{Stellm04}. The classification of the dynamos
above the transition is more difficult. 
Cattaneo and Hughes \cite{Cattan06} found convection driven
dynamos which present the characteristics expected
from dynamos in a chaotic flow with little or no helicity. These dynamos
generate magnetic fields at small length scales and with little mean field.
The distinction between generation at
large and small scales is pointless for the flows studied here because all flows
look similar to those visualized in ref. \cite{Stellm04}: They consist of long thin
vortices aligned with the rotation axis with little interior structure, so that
in horizontal planes, large and small scales are identical. The energy contained
in the mean magnetic field is less than 30 \% of the total magnetic energy in
all cases simulated here and this ratio smoothly decreases with increasing
magnetic Reynolds number. At $\mathrm{Rm}\, \mathrm{Ek}^{1/3}=13.5$, this ratio
is down to 7 \%, so that these features are not useful for identifying
the generation mechanism.

\begin{figure}
\includegraphics[width=8cm]{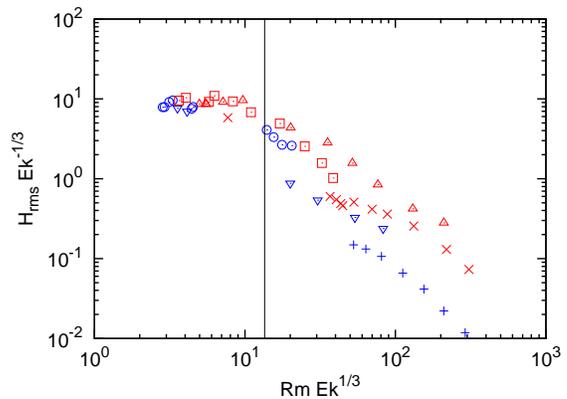}
\caption{(Color online)
$H_{\mathrm{rms}} \mathrm{Ek}^{-1/3}$
as a function of $\mathrm{Rm}\, \mathrm{Ek}^{1/3}$
for the successful dynamos of fig. \ref{fig:phase_diagram}. The
vertical line is located at $\mathrm{Rm}\, \mathrm{Ek}^{1/3}=13.5$.}
\label{fig:helicity}
\end{figure}

A more important hint is provided by helicity. Following refs.
\cite{Cattan06, Schmit10}, we define as helicity $\langle H \rangle$ the correlation
$\bm v \cdot \nabla \times \bm v /(|\bm v| \cdot |\nabla \times \bm v|)$
between velocity and vorticity averaged over horizontal planes and time, which
still depends on $z$. We obtain a single number $H_{\mathrm{rms}}$ as measure of
helicity from $H_{\mathrm{rms}}=\int_0^1 \langle H \rangle ^2 dz$.
As fig. \ref{fig:helicity} shows, $H_{\mathrm{rms}} \mathrm{Ek}^{-1/3}$ is
nearly constant below the transition and decreases above. The helicity is 
apparently held
constant by the magnetic field below the transition, because in nonmagnetic
convection, helicity continuously decreases with increasing Reynolds number
\cite{Schmit10}. The decrease of helicity above the transition does not prevent
magnetic fields from growing, which shows that helicity is not required for
dynamo action above the transition. We therefore arrive at a picture in which
there are two mechanisms responsible for dynamo action. One is based on helicity
and dominates below the transition, but it becomes less and less efficient at
high magnetic Reynolds numbers. The other is independent of helicity and takes
over at high magnetic Reynolds numbers. Field amplification by chaotic
stretching is intrinsically a mechanism requiring high magnetic Reynolds
numbers, which fits well with the fact that the transition occurs at
a certain value of $\mathrm{Rm}\, \mathrm{Ek}^{1/3}$.

Another type of transition is known from convection dynamos
in spheres, where the field can transit from a field dominated by its dipole
contribution to a field with a much smaller dipole moment. This transition
occurs at a Rossby number based on the size of the vortices in the flow of about
0.1 \cite{Christ06}, which in the present notation corresponds to 
$(\mathrm{Re}\, \mathrm{Ek})\mathrm{Ek}^{1/3}=0.052$. Nothing indicates a
transition as a function of this Rossby number in the present simulations. More
recent simulations in spherical shells \cite{Soderl12}, however, find the
transition when inertial and viscous forces become comparable.

\begin{figure}
\includegraphics[width=8cm]{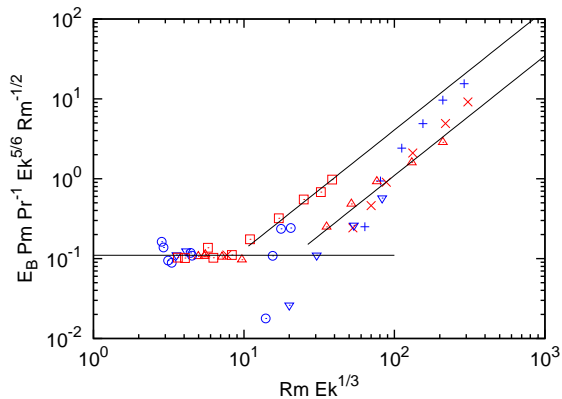}
\caption{(Color online)
$E_B \, \mathrm{Ek}^{5/6}\, \mathrm{Pm}/(\mathrm{Rm}^{1/2}\, \mathrm{Pr})$
as a function of $\mathrm{Rm}\, \mathrm{Ek}^{1/3}$
with the same symbols as in fig. \ref{fig:kinematic}. The horizontal line
marks the value of 0.11, and the other lines show
$(\mathrm{Rm}\, \mathrm{Ek}^{1/3})^{3/2}$.}
\label{fig:magnetostrophic}
\end{figure}

\begin{figure}
\includegraphics[width=8cm]{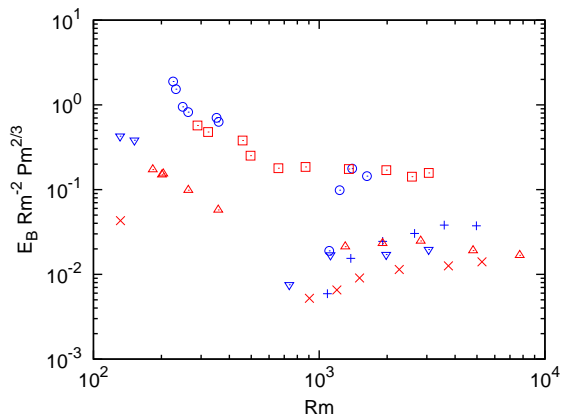}
\caption{(Color online)
Same data as in fig. \ref{fig:magnetostrophic} but plotted as
$E_B \, \mathrm{Rm}^{-2}\, \mathrm{Pm}^{2/3}$ versus
$\mathrm{Rm}$. Only the successful dynamos appear in figs.
\ref{fig:magnetostrophic} and \ref{fig:equipartition} since $E_B=0$ for the
others.}
\label{fig:equipartition}
\end{figure}

The scaling of the amplitude of the generated magnetic field in the
statistically stationary
state depends on whether the dynamo is below or above the transition.
Below the transition, the magnetic field strength obeys a scaling indicative of
a magnetostrophic balance, in which the Lorentz force is in equilibrium with the
Coriolis force. Some fraction of the Coriolis and Lorentz forces is
balanced by the pressure gradient, but it is impossible to compute that
fraction from the data actually saved during the simulations. We will equate naive estimates for
both forces, replacing the Coriolis and Lorentz terms in eq. (\ref{eq:NS_BQ}) by
$\mathrm{Re}\, \mathrm{Pr}/\mathrm{Ek}$ and $E_B/l_B$,
respectively, where $l_B$ stands for the length scale on which the magnetic
field typically varies. From simulations of pure convection \cite{Schmit09} we
know that the size of the convection columns stays close to the size they have
at the onset of convection throughout the parameter range of interest here, i.e.
close to $1.9 \cdot \mathrm{Ek}^{1/3}$. The magnetic Reynolds number based on
that size varies for the dynamos below the transition from 4.8 to 25. At these
Reynolds numbers, the magnetic field is not distributed uniformly but is
expelled from the core of the vortices and concentrates in boundary layers of
thickness proportional to $\mathrm{Ek}^{1/3} (\mathrm{Rm}\,
\mathrm{Ek}^{1/3})^{-1/2}$. Using this expression for $l_B$, the assumption of
magnetostrophic balance leads to 
$E_B \, \mathrm{Ek}^{5/6}\, \mathrm{Pm}/(\mathrm{Rm}^{1/2}\, \mathrm{Pr}) =
const.$. It is verified in fig. \ref{fig:magnetostrophic} that the left hand
side of this equation is $0.11 \pm 0.03$ for all dynamos below the transition,
the largest deviations being due to the simulations at low $\mathrm{Ek}$ with
widely fluctuating energies typical for this parameter range 
\cite{Stellm04} and which require long runs
before accurate values of $E_B$ and $\mathrm{Rm}$ are obtained.

As suggested by fig. \ref{fig:equipartition} and
the asymptotes drawn into fig. \ref{fig:magnetostrophic}, the
magnetic field obeys a scaling close to $E_B \propto \mathrm{Rm}^2$ above the
transition. Such a behavior arises if there is equipartition between kinetic and
magnetic energies, or if there is a balance between advection and Lorentz terms
and if the same length scale is used to estimate these two terms in eq.
(\ref{eq:NS_BQ}). None of this is apparently exactly realized in the present simulations
because a nontrivial dependence on $\mathrm{Ek}$ and $\mathrm{Pm}$ remains.


\acknowledgments
This work was supported by the Deutsche Forschungsgemeinschaft (DFG).


\begin{thebibliography}{15}
\expandafter\ifx\csname natexlab\endcsname\relax\def\natexlab#1{#1}\fi
\expandafter\ifx\csname bibnamefont\endcsname\relax
  \def\bibnamefont#1{#1}\fi
\expandafter\ifx\csname bibfnamefont\endcsname\relax
  \def\bibfnamefont#1{#1}\fi
\expandafter\ifx\csname citenamefont\endcsname\relax
  \def\citenamefont#1{#1}\fi
\expandafter\ifx\csname url\endcsname\relax
  \def\url#1{\texttt{#1}}\fi
\expandafter\ifx\csname urlprefix\endcsname\relax\def\urlprefix{URL }\fi
\providecommand{\bibinfo}[2]{#2}
\providecommand{\eprint}[2][]{\url{#2}}

\bibitem[{\citenamefont{Jones and Roberts}(2000)}]{Jones00}
\bibinfo{author}{\bibfnamefont{C.}~\bibnamefont{Jones}} \bibnamefont{and}
  \bibinfo{author}{\bibfnamefont{P.}~\bibnamefont{Roberts}},
  \bibinfo{journal}{J. Fluid Mech.} \textbf{\bibinfo{volume}{404}},
  \bibinfo{pages}{311} (\bibinfo{year}{2000}).

\bibitem[{\citenamefont{Rotvig and Jones}(2002)}]{Rotvig02}
\bibinfo{author}{\bibfnamefont{J.}~\bibnamefont{Rotvig}} \bibnamefont{and}
  \bibinfo{author}{\bibfnamefont{C.}~\bibnamefont{Jones}},
  \bibinfo{journal}{Phys. Rev. E} \textbf{\bibinfo{volume}{66}},
  \bibinfo{pages}{056308} (\bibinfo{year}{2002}).

\bibitem[{\citenamefont{Stellmach and Hansen}(2004)}]{Stellm04}
\bibinfo{author}{\bibfnamefont{S.}~\bibnamefont{Stellmach}} \bibnamefont{and}
  \bibinfo{author}{\bibfnamefont{U.}~\bibnamefont{Hansen}},
  \bibinfo{journal}{Phys. Rev. E} \textbf{\bibinfo{volume}{70}},
  \bibinfo{pages}{056312} (\bibinfo{year}{2004}).

\bibitem[{\citenamefont{Cattaneo and Hughes}(2006)}]{Cattan06}
\bibinfo{author}{\bibfnamefont{F.}~\bibnamefont{Cattaneo}} \bibnamefont{and}
  \bibinfo{author}{\bibfnamefont{D.}~\bibnamefont{Hughes}},
  \bibinfo{journal}{J. Fluid Mech.} \textbf{\bibinfo{volume}{553}},
  \bibinfo{pages}{401} (\bibinfo{year}{2006}).

\bibitem[{\citenamefont{Schmitz and Tilgner}(2009)}]{Schmit09}
\bibinfo{author}{\bibfnamefont{S.}~\bibnamefont{Schmitz}} \bibnamefont{and}
  \bibinfo{author}{\bibfnamefont{A.}~\bibnamefont{Tilgner}},
  \bibinfo{journal}{Phys. Rev. E} \textbf{\bibinfo{volume}{80}},
  \bibinfo{pages}{015305(R)} (\bibinfo{year}{2009}).

\bibitem[{\citenamefont{King et~al.}(2009)\citenamefont{King, Stellmach, Noir,
  Hansen, and Aurnou}}]{King09}
\bibinfo{author}{\bibfnamefont{E.~M.} \bibnamefont{King}},
  \bibinfo{author}{\bibfnamefont{S.}~\bibnamefont{Stellmach}},
  \bibinfo{author}{\bibfnamefont{J.}~\bibnamefont{Noir}},
  \bibinfo{author}{\bibfnamefont{U.}~\bibnamefont{Hansen}}, \bibnamefont{and}
  \bibinfo{author}{\bibfnamefont{J.~M.} \bibnamefont{Aurnou}},
  \bibinfo{journal}{Nature} \textbf{\bibinfo{volume}{457}},
  \bibinfo{pages}{301} (\bibinfo{year}{2009}).

\bibitem[{\citenamefont{Chorin}(1967)}]{Chorin67}
\bibinfo{author}{\bibfnamefont{A.}~\bibnamefont{Chorin}}, \bibinfo{journal}{J.
  Comp. Phys.} \textbf{\bibinfo{volume}{2}}, \bibinfo{pages}{12}
  (\bibinfo{year}{1967}).

\bibitem[{\citenamefont{Tanriverdi and Tilgner}(2011)}]{Tanriv11}
\bibinfo{author}{\bibfnamefont{V.}~\bibnamefont{Tanriverdi}} \bibnamefont{and}
  \bibinfo{author}{\bibfnamefont{A.}~\bibnamefont{Tilgner}},
  \bibinfo{journal}{New Journal of Physics} \textbf{\bibinfo{volume}{13}},
  \bibinfo{pages}{033019} (\bibinfo{year}{2011}).
\bibitem[{\citenamefont{Tilgner}(2011)}]{Tilgne11}
\bibinfo{author}{\bibfnamefont{A.}~\bibnamefont{Tilgner}},
  \bibinfo{journal}{Phys. Rev. E} \textbf{\bibinfo{volume}{84}},
  \bibinfo{pages}{026323} (\bibinfo{year}{2011}).

\bibitem[{\citenamefont{Soward}(1974)}]{Soward74}
\bibinfo{author}{\bibfnamefont{A.}~\bibnamefont{Soward}},
  \bibinfo{journal}{Phil. Trans. R. Soc. Lond. A}
  \textbf{\bibinfo{volume}{275}}, \bibinfo{pages}{611} (\bibinfo{year}{1974}).

\bibitem[{\citenamefont{Schmitz and Tilgner}(2010)}]{Schmit10}
\bibinfo{author}{\bibfnamefont{S.}~\bibnamefont{Schmitz}} \bibnamefont{and}
  \bibinfo{author}{\bibfnamefont{A.}~\bibnamefont{Tilgner}},
  \bibinfo{journal}{Geophys. Astrophys. Fluid Dynam.}
  \textbf{\bibinfo{volume}{104}}, \bibinfo{pages}{481} (\bibinfo{year}{2010}).

\bibitem[{\citenamefont{Stevens et~al.}(2010)\citenamefont{Stevens, Clercx, and
  Lohse}}]{Steven10}
\bibinfo{author}{\bibfnamefont{R.}~\bibnamefont{Stevens}},
  \bibinfo{author}{\bibfnamefont{H.~J.~H.} \bibnamefont{Clercx}},
  \bibnamefont{and} \bibinfo{author}{\bibfnamefont{D.}~\bibnamefont{Lohse}},
  \bibinfo{journal}{New J. Phys.} \textbf{\bibinfo{volume}{12}},
  \bibinfo{pages}{075005} (\bibinfo{year}{2010}).
\bibitem[{\citenamefont{Kunnen et~al.}(2006)\citenamefont{Kunnen, Clercx, and
  Geurts}}]{Kunnen06}
\bibinfo{author}{\bibfnamefont{R.~P.~J.} \bibnamefont{Kunnen}},
  \bibinfo{author}{\bibfnamefont{H.~J.~H.} \bibnamefont{Clercx}},
  \bibnamefont{and} \bibinfo{author}{\bibfnamefont{B.~J.}
  \bibnamefont{Geurts}}, \bibinfo{journal}{Phys. Rev. E}
  \textbf{\bibinfo{volume}{74}}, \bibinfo{pages}{056306}
  (\bibinfo{year}{2006}).

\bibitem[{\citenamefont{Christensen and Aubert}(2006)}]{Christ06}
\bibinfo{author}{\bibfnamefont{U.}~\bibnamefont{Christensen}} \bibnamefont{and}
  \bibinfo{author}{\bibfnamefont{J.}~\bibnamefont{Aubert}},
  \bibinfo{journal}{Geophys. J. Int.} \textbf{\bibinfo{volume}{166}},
  \bibinfo{pages}{97} (\bibinfo{year}{2006}).

\bibitem[{\citenamefont{Soderlund et~al.}(2012)\citenamefont{Soderlund, King,
  and Aurnou}}]{Soderl12}
\bibinfo{author}{\bibfnamefont{K.}~\bibnamefont{Soderlund}},
  \bibinfo{author}{\bibfnamefont{E.}~\bibnamefont{King}}, \bibnamefont{and}
  \bibinfo{author}{\bibfnamefont{J.}~\bibnamefont{Aurnou}},
  \bibinfo{journal}{Earth Planet. Sci. Lett.}
  \textbf{\bibinfo{volume}{333-334}}, \bibinfo{pages}{9}
  (\bibinfo{year}{2012}).

\end{thebibliography}

\end{document}